\documentclass[a4paper,11pt]{article}
\usepackage{pos}

\usepackage{graphicx}
\usepackage{dcolumn}
\usepackage{bm,graphicx,hyperref,subfigure}
\usepackage{float,amsfonts,graphicx}
\usepackage{color}
\usepackage{tikz}
\usepackage{adjustbox}
\usetikzlibrary{decorations.pathreplacing}
\usetikzlibrary{arrows.meta}
\relax

\DeclareMathOperator{\Tr}{\mathrm{Tr}}

\DeclareMathOperator{\dd}{{\rm{d}}}

\newcommand{\eq}[1]{\begin{align}\label{#1}}
\newcommand{\en}{\end{align}}
\newcommand{\eqar}[1]{\begin{align}\label{#1}}
\newcommand{\enar}{\end{align}}

\usepackage{subcaption}

\newcommand{\ie}{i.e. }

\title{Duality and entanglement in lattice gauge theories}

\author*[a]{Andrea Bulgarelli}
\author[a]{Marco Panero}

\affiliation[a]{Physics Department, University of Turin \& INFN, Turin unit\\Via Pietro Giuria 1, I-10125 Turin, Italy}

\emailAdd{andrea.bulgarelli@unito.it}

\abstract{The study of entanglement in quantum field theories provides insight into universal properties which are typically challenging to extract by means of local observables. However, calculations of quantities related to entanglement in gauge theories are limited by ambiguities that stem from the non-factorizability of the Hilbert space. On the other hand, (2+1)-dimensional lattice gauge theories are known to admit a dual description in terms of spin models, for which the replica trick and Rényi entropies are well defined. In this contribution, we exploit Kramers-Wannier duality to perform a numerical study of the entropic c-function of the (2+1)-dimensional $\mathbb{Z}_2$ gauge theory in the continuum limit. Our results are compared with analytical predictions from holographic models.}

\FullConference{The 41st International Symposium on Lattice Field Theory (Lattice 2024)\\
July 28th - August 3rd, 2024\\
University of Liverpool}

\begin{document}
\maketitle

\section{Introduction}

Gauge theories represent a central paradigm in physics. One one hand, the fundamental interactions in Nature can be described in terms of gauge theories, which constitute the backbones of the Standard Model of particle physics. Also, in condensed matter systems, gauge theories are realized as low-energy descriptions of different phases of matter. As such, the study of quantum phenomena emerging as a consequence of gauge symmetry is relevant for many areas of physics.

Among the many characteristic phenomena of quantum systems, one of the most ubiquitous is entanglement, \ie the existence of quantum correlations that cannot be described in classical terms~\cite{Horodecki:2009zz}. Apart from being a practical resource in quantum simulations~\cite{nielsen2000quantum}, the study of entanglement in many-body quantum systems finds applications ranging from quantum phase transitions~\cite{Vidal:2002rm} to quantum gravity and holography~\cite{Ryu:2006bv, VanRaamsdonk:2010pw}.

Such correlations become evident as one tries to give a description of a portion of a quantum system alone; specifically, given a Hilbert space which admits a factorization $\mathcal{H} = \mathcal{H}_A\otimes\mathcal{H}_B$, where $A$ and $B$ are two complementary subsystems, one can define a reduced density matrix encoding information on the subsystem $A$ alone by tracing the density matrix $\rho$ of the system over a complete basis of $B$, $\rho_A = \Tr_B\rho$. If $\rho$ describes a pure state, then the Rényi entropies,
\begin{align}
  S_n = \frac{1}{1-n}\log\Tr\rho_A^n,
  \label{definition_Renyi_entropy}
\end{align}
measure the amount of bipartite entanglement between $A$ and $B$, being entanglement monotones~\cite{Vidal:1998re}. The entanglement entropy $S = -\Tr(\rho_A\log\rho_A)$ is recovered as the $n\rightarrow 1$ limit of the Rényi entropy of order $n$.

Given the central r\^ ole of gauge theories in physics and, at the same time, the importance of entanglement in the study of quantum phenomena, it comes as a problem the fact that defining entanglement in a gauge theory is a non-trivial matter. The presence of local symmetries, and the consequent Gau\ss 's law, imply that the gauge invariant portion of the Hilbert space does not admit a tensor product decomposition in subspaces with only gauge invariant states~\cite{Buividovich:2008gq, Donnelly:2011hn, Casini:2013rba, Radicevic:2014kqa, Ghosh:2015iwa, Aoki:2015bsa, Radicevic:2015sza, Soni:2015yga, Lin:2018bud}. For the purposes of this work, the main consequence is that calculations of entanglement measures in gauge theories, e.g. with Monte Carlo simulations, are plagued with ambiguities; this fact limits the possibility of investigating highly interesting phenomena through the lens of entanglement.

One of such phenomena is confinement. Following the paper by Klebanov et al.~\cite{Klebanov:2007ws}, there have been different works using Monte Carlo simulations to study the behavior of the Rényi entropy and related quantities in gauge theories~\cite{Buividovich:2008kq, Itou:2015cyu, Rabenstein:2018bri, Jokela:2023rba, Bulgarelli:2024onj, Amorosso:2024leg}. All of these studies exploit a discretized version of the replica trick~\cite{Calabrese:2004eu}, where the Rényi entropy is expressed in terms of a ratio of partition functions
\begin{align}
  \Tr\rho_A^n = \frac{Z_n}{Z^n},
  \label{ratio_of_partition_functions_replica_trick}
\end{align}
where $Z^n$ denotes the partition function of $n$ independent copies of a given system, while $Z_n$ represents the same theory on a Riemann surface, obtained by gluing together the copies through a cut, corresponding to the subsystem $A$ at fixed Euclidean time. However, due to the aforementioned ambiguity in the factorization of the Hilbert space, it is unclear how a lattice version of the Riemann surface should be constructed, and in particular how different replicas should be glued together.

In~\cite{Bulgarelli:2024onj}, we addressed this problem by providing an unambiguous way to compute entanglement measures in gauge theories which admit a  Kramers-Wannier dual~\cite{Kramers:1941kn, Kramers:1941zz}. This, together with the efficient algorithm developed in~\cite{Bulgarelli:2023ofi}, allowed us to perform a high precision test of the conjectures of~\cite{Klebanov:2007ws}, finding good agreement with the theoretical predictions.

\section{Entanglement in gauge theories}

In gauge theories, the lack of a natural way to factorize the gauge invariant Hilbert space leads to different approaches to study entanglement, among which two are broadly used. In the extended Hilbert space approach~\cite{Buividovich:2008gq} one locally enlarges the gauge invariant Hilbert space, including a minimal amount of states violating Gau\ss 's law to ensure factorizability. Here, instead, we review the second approach~\cite{Casini:2013rba}, which focuses on the algebra of gauge invariant operators acting on the Hilbert space of the theory.

In an Abelian\footnote{For non-Abelian theories the analysis is essentially the same with some minor modifications.} lattice gauge theory the algebra of operators $\mathcal{A}$ is generated by the field (or coordinate) operators $U^r_l$, where $l$ labels the link and $r$ the one-dimensional representation of the gauge group, and by the electric-field (or momentum) operator $L_l^g$, associated with the element of the gauge group $g$. As the field operators change under gauge transformations, one can further define a subset $\mathcal{A}_G$ of gauge invariant operators which includes Wilson loops $W^r_{\Gamma} = \prod_{l\in\Gamma}U^r_l$, where $\Gamma$ is a closed loop, as well as the electric-field operators. The crucial idea of the operator approach is that we can then identify a spatial subsystem $A$ as the subalgebra $\mathcal{A}_G(A) \in \mathcal{A}_G$ acting on $A$. The reduced density matrix is then the unique operator in $\mathcal{A}_G(A)$ satisfying $\Tr(\mathcal{O}\rho_A) = \Tr(\mathcal{O}\rho)$ $\forall \mathcal{O} \in \mathcal{A}_G(A)$.

The reduced density matrix obtained in this way is by construction gauge invariant, but it strongly depends on the choice of the subalgebra, which is arbitrary to some extent. Indeed, different choices of boundary operators lead to different centers of the algebra, i.e. the subset $\mathcal{Z}\subset\mathcal{A}_G(A)$ commuting with all the other elements of the algebra. In particular, in the so called electric-center algebra, one includes all the electric operators at the boundary, while excluding all of them leads to the magnetic-center choice~\cite{Casini:2013rba}.

As a consequence of the presence of a non-trivial center, all the operators in $\mathcal{A}_G(A)$, including $\rho_A$, are block diagonal in a basis that diagonalizes $\mathcal{Z}$, with different blocks identifying different superselection sectors of the theory. In particular
\begin{align}
\rho_A = \begin{pmatrix}
  p_1\rho_1 & 0 & \dots & 0\\
  0 & p_2\rho_2 & \dots & 0 \\
  \vdots & \vdots & \ddots & \vdots \\
  0 & 0 & \dots & p_m\rho_m
\end{pmatrix},
\end{align}
where $p_k$ is the probability distribution of the superselection sectors; the entanglement entropy can then be decomposed as
\begin{align}
  S = -\sum_k p_k\log p_k + \sum_k p_k S(\rho_k),
  \label{entanglement_entropy_in_a_gauge_theory}
\end{align}
and a similar equation holds for the Rényi entropy. In the previous expression, the first term, which is the Shannon entropy of the probability distribution $p_k$, is sensitive to the choice of the center, which is a purely ultraviolet (UV) piece of information, which in principle should not affect the infrared (IR) physics. It was indeed found that quantities with a well defined continuum limit, such as the mutual information~\cite{Casini:2013rba, Casini:2014aia}, converge toward the same IR value regardless of the choice of $\mathcal{Z}$.

It is therefore possible to use entanglement related quantities to unambiguously investigate the IR physics of gauge theories. In doing that, however, as the explicit structure of the superselection sectors as well as the calculation of the two terms of \eqref{entanglement_entropy_in_a_gauge_theory} separately is beyond the reach of typical numerical methods, one has to rely on the replica trick and lattice simulations. The transition from the choice of an algebra in the Hamiltonian formulation to the construction of a replica space in the Lagrangian picture is non-trivial~\cite{Lin:2018bud}, especially as one further seeks to discretize the replica space to perform numerical simulations.

The approach we followed in~\cite{Bulgarelli:2024onj} is based on the fact that, for spin models with a factorizable Hilbert space, no ambiguities arise in the construction of a discretized replica space. At the same time, there exist spin models that can be exactly mapped to gauge theories, providing a way to bypass the need of a direct definition of the replica geometry in the gauge theory. Here, the map we are interested in is the Kramers-Wannier duality.

\section{Kramers-Wannier duality}

The Kramers-Wannier duality~\cite{Kramers:1941kn, Kramers:1941zz} is a transformation between two different lattice theories preserving the partition function, up to numerical constants (a detailed review can be found in~\cite{Savit:1980}). In what follows, we focus on dualities in $D = 2+1$ dimensions for isotropic cubic lattices, with an infinite extent in all directions. For concreteness, we focus on the Ising model, though the discussion can be generalized to other spin models with global Abelian symmetries.

By means of a discrete Fourier transform, the partition function of the $3D$ Ising model
\begin{align}
  Z(\beta) = \sum_{\{\sigma\}}\exp\left(\beta\sum_i\sum_{\mu=0}^2 \sigma_i\sigma_{i+\hat{\mu}}\right),
\end{align}
can be mapped, up to numerical factors, to the partition function of the $\mathbb{Z}_2$ lattice gauge theory
\begin{align}
  Z^{\text{gauge}}(\beta^*) = \sum_{\{U\}}\exp\left(\beta^*\sum_{i}\sum_{\mu,\nu=0}^2 U_{i,\mu}U_{i+\hat{\mu},\nu}U^\dagger_{i+\hat{\nu},\mu}U^\dagger_{i,\nu}\right),
\end{align}
where $U\in\{\pm 1\}$ and the coupling $\beta^*$ is connected to $\beta$ through the relation
\begin{align}
  \beta^* = -\frac{1}{2}\log\tanh\beta.
\end{align}
In~\cite{Bulgarelli:2024onj} we made use of this relation to map the ratio of partition functions~\eqref{ratio_of_partition_functions_replica_trick} from the Ising model to the $\mathbb{Z}_2$ gauge theory, finding
\begin{align}
  \frac{Z_n}{Z^n}|_{\text{Ising}} \propto \frac{Z_n}{Z^n}|_{\text{gauge}}.
\end{align}
As the previous ratio directly enters the calculation of the Rényi entropy, it turns out that the entropy itself is not mapped through the duality transformation. In particular, the entropy of the gauge model is larger than the entropy of the spin model, consistently with other studies in literature~\cite{Tagliacozzo:2010vk}. At the same time, however, both Rényi and entanglement entropies depend on UV details, such as the lattice discretization and the choice of the algebra; UV-finite quantities defined out of these entropies are not only expected to converge towards the same IR limit, regardless of the UV details, but also to be preserved under the Kramers-Wannier duality~\cite{Moitra:2018lxn}.

The UV-finite quantity we analyze here is the entropic c-function
\begin{align}
    C_n = \frac{l^{D-1}}{|\partial A|}\frac{\partial S_n}{\partial l},
\end{align}
where $D$ is the number of spacetime dimensions, $|\partial A|$ the area of the entangling surface and $l$ the linear size of the subsystem $A$. Remarkably, this quantity is known to be proportional to the number of effective degrees of freedom of a given quantum field theory~\cite{Casini:2006es, Zamolodchikov:1986gt}. We will further discuss this aspect in the following section. By means of the replica trick, the entropic c-function also admits the following expression
\begin{align}
    C_n = \frac{l^{D-1}}{|\partial A|}\frac{1}{n-1}\lim_{a\rightarrow 0}\frac{1}{a}\log\frac{Z_n(l)}{Z_n(l+a)}.
    \label{entropic_c-function_replica_trick}
\end{align}
On the lattice, the derivative with respect to $l$ can be approximated by a finite difference and $a$ becomes the smallest length scale that can be resolved, namely the lattice spacing. As discussed in detail in~\cite{Bulgarelli:2024onj}, one can show that the Karamers-Wannier duality preserves the entropic c-function, \ie
\begin{align}
  C_n^{\text{Ising}} = C_n^{\text{gauge}}.
  \label{equality_of_cfunction_through_Kramers-Wannier}
\end{align}
This equality can be used in two ways. On one hand, starting from the well defined replica space of the spin model, one can unambiguously derive the dual replica geometry. This is discussed in detail in~\cite{Bulgarelli:2024onj}, where we analyzed the dual geometry of a number of theories, including the $\mathbb{Z}_2$ and the $U(1)$ gauge theories. Here, we focus on a complementary approach: by simulating the replica space of the Ising model, we compute the left-hand side of ~\eqref{equality_of_cfunction_through_Kramers-Wannier} by means of Monte Carlo simulations, and use the result to study the entropic c-function of the dual gauge theory.

\section{Monte Carlo simulations}

\begin{figure*}[t]
\centerline{\includegraphics[height=0.4\textwidth]{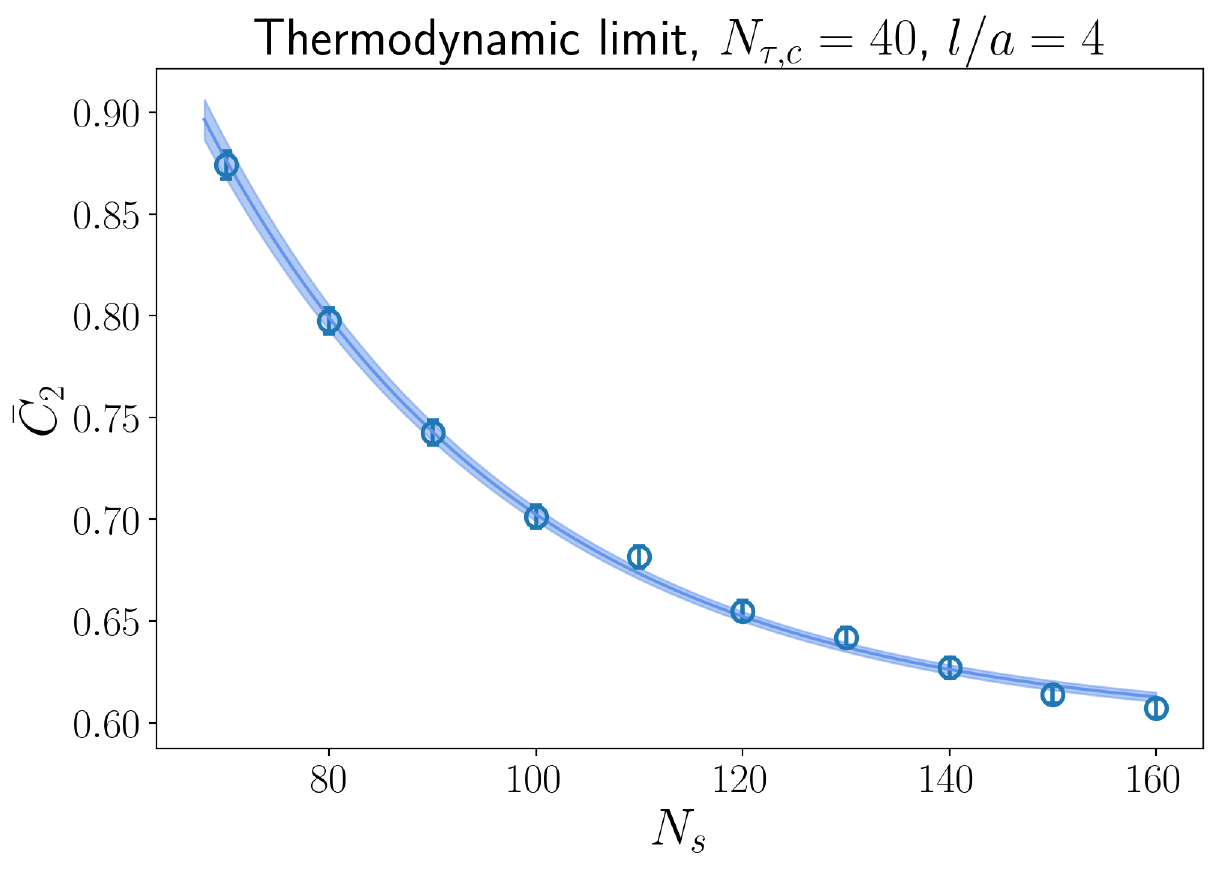} \hfill \includegraphics[height=0.4\textwidth]{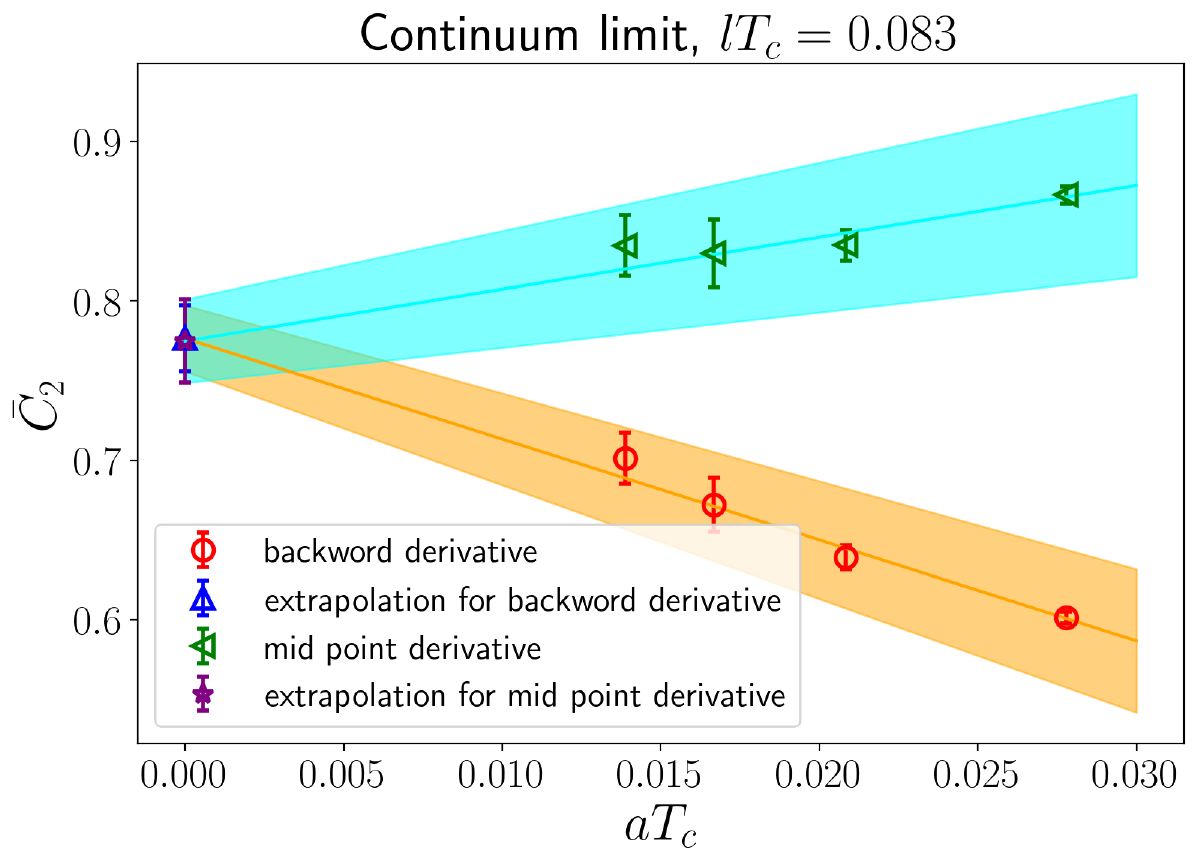}}
\caption{An example of a thermodynamic (left panel) and a continuum (right panel) limits.}
\label{fig:thermodynamic_and_continuum_limit}
\end{figure*}

In this section we discuss a Monte Carlo study of the ground-state entropic c-function of the $(2+1)$-dimensional $\mathbb{Z}_2$ lattice gauge theory, exploiting duality. The target quantity of the simulation is the ratio $Z_n(l)/Z_n(l+a)$~\eqref{entropic_c-function_replica_trick}, therefore an efficient algorithm to compute ratios of partition functions is needed. In recent years there has been significant progress in non-equilibrium Monte Carlo simulations, which, being based on the Jarzynski's theorem~\cite{Jarzynski:1996oqb}, yield very precise estimates of ratios of partition functions~\cite{Caselle:2016wsw, Caselle:2018kap, Francesconi:2020fgi, Caselle:2022acb, Bonanno:2024udh, Caselle:2024ent, Bulgarelli:2024cqc}; in particular, we used the algorithm introduced in~\cite{Bulgarelli:2023ofi} as a generalization of~\cite{Alba:2016bcp}.

We simulated a two-replica system to extract the entropic c-function associated to the second Rényi entropy; in particular, we chose the subsystem $A$ to be a slab, maximally extended in one of the two spatial directions, and with thickness $l/a$ in the other one. In all the simulations we fixed the length of the Euclidean time direction to be at least $10 \, N_{\tau,c}$, where $N_{\tau,c}$ is the critical length of the deconfinement transition for the $\mathbb{Z}_2$ gauge theory. In this regime, thermal fluctuations are suppressed and $C_2$ receives contributions only from genuinely quantum fluctuations. For the scale setting we followed ref.~\cite{Caselle:1995wn}.

We focused on the confining regime of the $\mathbb{Z}_2$ gauge theory, which is mapped to the broken-symmetry phase of the Ising model, $\beta > \beta_c$. By evaluating $C_2$ for a range of volumes and couplings, we were able to perform the first thermodynamic and continuum extrapolation of such quantity in a confining gauge theory in literature, see Fig.~\ref{fig:thermodynamic_and_continuum_limit}. Being our results free of lattice artifacts, we can then make comparison with analytical predictions.

In~\cite{Klebanov:2007ws}, the authors focused on the entanglement entropy of confining quantum field theories with a holographic dual. The calculation of the entanglement entropy in a slab geometry, as the one we used for our simulations, leads to the following results. As the thickness of the slab is small compared to typical length-scales of the theory, the entropic c-function scales as $N_c^2$, where $N_c$ is the number of colors. In a QCD-like theory, this corresponds to the number of gluons in the theory. For larger slabs the entropic c-function undergoes a sharp transition, becoming insensitive to the number of colors of the theory, as the long-distance spectrum of the theory is indeed made of colorless excitations, e.g. glueballs or hadrons. The behavior of the c-function is indeed consistent with its relation with the effective number of degrees of freedom of a theory. Furthermore, this motivated the authors of~\cite{Klebanov:2007ws} to conjecture that entanglement can be used to probe confinement.

In non-holographic theories, such as $SU(N)$ gauge theories, analytic calculations are unfeasible; nonetheless in~\cite{Klebanov:2007ws} it was conjectured that such theories can display an exponential decay for large enough distances, rather than a sharp transition. One can indeed approximate a gas of glueballs as weakly interacting bosons. In $1+1$ dimensions the entropic c-function of a free scalar is known analytically~\cite{Casini:2005zv}, and one can obtain the same result in higher dimension by means of dimensional reduction. In $2+1$ dimensions, in particular, one obtains the following result, up to a multiplicative constant
\begin{align}
  C_2 = l\int \dd \mathbf{k} \exp\left(-2\sqrt{m^2 + \mathbf{k}^2}l\right),
  \label{conjecture_cfunction_non_holographic}
\end{align}
where $l$ is the length of the slab and $m$ the mass of the scalar.

\begin{figure}[t]
\centering
\includegraphics[width=0.7\textwidth]{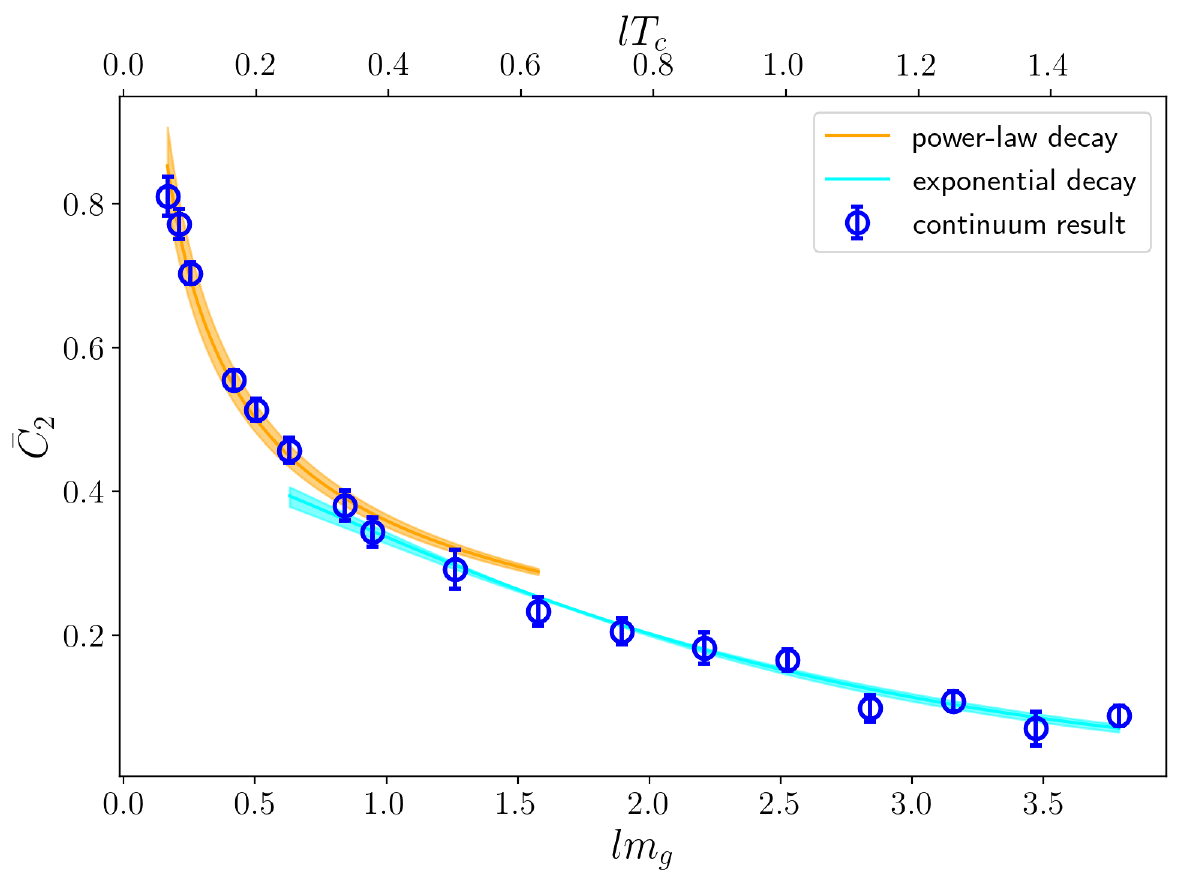}
\caption{Thermodynamic and continuum extrapolation of $C_2$ in the $(2+1)$-dimensional $\mathbb{Z}_2$ gauge theory. Data are compared with two models, a power-law decay at short distances (compared to the inverse mass-gap), and an exponential decay at large distances.}
\label{fig:cfunction_Z2_gauge_theory}
\end{figure}

In Fig.~\ref{fig:cfunction_Z2_gauge_theory} the results of the thermodynamic and continuum extrapolation of the entropic c-function in the $(2+1)$-dimensional $\mathbb{Z}_2$ gauge theory are displayed. $C_2$ is normalized such that
\begin{align}
  \bar{C}_2(l) = \frac{C_2(l)}{C_2^{\text{CFT}}} \qquad C_2^{\text{CFT}} = C_2(0).
\end{align}
The value of $C_2^{\text{CFT}}$ has been determined in~\cite{Kulchytskyy:2019hft}. Here, we use as a reference scale the inverse mass-gap of the theory $m_{g}$, whose value has been taken from~\cite{Agostini:1996xy}. At small length scales we modeled our data with a power-law decay
\begin{align}
  f(lm_g;B,c) = B/(lm_g)^c,
  \label{fit_function_polynomal}
\end{align}
obtaining as best fit estimates of the parameters $B=0.360(9)$ and $c=0.48(2)$, with a reduced chi-squared $\chi^2_{\text{red}} = 1.02$ (orange curve in Fig.~\ref{fig:cfunction_Z2_gauge_theory}). At large length scales, we fitted the data with the function
\begin{align}
  f(lm_g; A, \alpha) = A\,lm_g\int \dd \mathbf{k} \exp\left(-2\alpha\sqrt{1 + \mathbf{k}^2}lm_g\right),
  \label{fit_function_exponential}
\end{align}
finding $A = 0.33(3)$ and $\alpha = 0.360(19)$, with $\chi^2_{\text{red}} = 0.84$ (cyan curve in Fig.~\ref{fig:cfunction_Z2_gauge_theory}).

The following comments are now in order. First, this study~\cite{Bulgarelli:2024onj} is the first numerical confirmation of the prediction~\eqref{conjecture_cfunction_non_holographic}, as our data are perfectly described by such functional form. Second, we are also able to determine which length scale of the theory is responsible for the transition from a power-law to an exponential decay: indeed, both the approximations~\eqref{fit_function_polynomal} and~\eqref{fit_function_exponential} break down around $lm_g = 1$. Notice that a very similar behavior has been pointed out in a study of other entanglement measures in the massless and massive Schwinger model~\cite{Florio:2023mzk}.

\section{Conclusions}

In this contribution we discussed the proposals we made in~\citep{Bulgarelli:2024onj} of using Kramers-Wannier duality as a tool for unambiguous studies of the entanglement content of Abelian gauge theories. In particular, one can either derive the replica geometry of the gauge theory by dualizing the spin-model lattice, or perform simulations of the entropic c-function of the spin model to study the same quantity on the gauge-theory side of the duality. In this contribution we discussed the second approach and we illustrated the Monte Carlo simulation we performed to extract the continuum entropic c-function of the $\mathbb{Z}_2$-gauge theory in $2+1$ dimensions. Our analysis is the first numerical confirmation of the prediction done in~\cite{Klebanov:2007ws} in a $(2+1)$-dimensional, non-holographic theory.

The same study can be done in any Abelian gauge theory admitting a dual description in terms of a spin model, such as the $U(1)$ gauge theory. For the non-Abelian case, even though duality transformations are known, they are much more involved, and a better theoretical understanding is needed. Also, moving to gauge theories with continuous groups might require more powerful algorithms to estimate ratios of partition functions, such as flow-based techniques~\cite{Bulgarelli:2024yrz}. We plan to address these research directions in future works.

\subsection*{Acknowledgments}
\noindent
This work of has been partially supported by the Spoke 1 ``FutureHPC \& BigData'' of the Italian Research Center on High-Performance Computing, Big Data and Quantum Computing (ICSC) funded by MUR (M4C2-19) -- Next Generation EU (NGEU), by the Italian PRIN ``Progetti di Ricerca di Rilevante Interesse Nazionale -- Bando 2022'', prot. 2022TJFCYB, and by the ``Simons Collaboration on Confinement and QCD Strings'' funded by the Simons Foundation. The simulations were run on CINECA computers. We acknowledge support from the SFT Scientific Initiative of the Italian Nuclear Physics Institute (INFN).

\bibliographystyle{JHEPjus}
\bibliography{pos2024}

\end{document}